\documentclass[preprint,12pt]{elsarticle}

\usepackage{amssymb}
\usepackage{algorithm}
\usepackage{algpseudocode}
\usepackage{amsmath}
\usepackage{array}
\usepackage{tikz}
\usepackage{listings}
\usepackage{xcolor}
\usepackage{subfigure}
\usetikzlibrary{positioning, shapes.geometric, arrows.meta}
\usepackage{booktabs} 
\usepackage{amsmath}
\usepackage{algorithm}
\usepackage{algpseudocode}

\journal{}

\begin{document}

\begin{frontmatter}

\title{AK-SLRL: Adaptive Krylov Subspace Exploration Using Single-Life Reinforcement Learning for Sparse Linear System}

\author[a,b]{Hadi Keramati}
\author[a]{Feridun Hamdullahpur}
\affiliation[a]{organization={University of Waterloo},
            addressline={200 University Avenue}, 
            city={Waterloo},
            postcode={N2L 3G1}, 
            state={Ontario},
            country={Canada}}

\affiliation[b]{organization={Magnative AI},
            addressline={215 Fort York Blvd}, 
            city={Toronto},
            postcode={M5V 4A2}, 
            state={Ontario},
            country={Canada}}

\date{}
\begin{abstract}
This paper presents a single-life reinforcement learning (SLRL) approach to adaptively select the dimension of the Krylov subspace during the generalized minimal residual (GMRES) iteration. GMRES is an iterative algorithm for solving large and sparse linear systems of equations in the form of \(Ax = b\) which are mainly derived from partial differential equations (PDEs). The proposed framework uses RL to adjust the Krylov subspace dimension (m) in the GMRES (m) algorithm. This research demonstrates that altering the dimension of the Krylov subspace in an online setup using SLRL can accelerate the convergence of the GMRES algorithm by more than an order of magnitude. A comparison of different matrix sizes and sparsity levels is performed to demonstrate the effectiveness of adaptive Krylov subspace exploration using single-life RL (AK-SLRL). We compare AK-SLRL with constant-restart GMRES by applying the highest restart value used in AK-SLRL to the GMRES method. The results show that using an adjustable restart parameter with single-life soft-actor critic (SLSAC) and an experience replay buffer sized to half the matrix dimension converges significantly faster than the constant restart GMRES with higher values. Higher values of the restart parameter are equivalent to a higher number of Arnoldi iterations to construct an orthonormal basis for the Krylov subspace $ K_m(A, r_0) $. This process includes constructing $m$ orthonormal vectors and updating the Hessenberg matrix $H$. Therefore, lower values of $m$ result in reduced computation needed in GMRES minimization to solve the least squares problem in the smaller Hessenberg matrix. The robustness of the result is validated through a wide range of matrix dimensions and sparsity. This paper contributes to the series of RL combinations with numerical solvers to achieve accelerated scientific computing.

\end{abstract}



\begin{keyword}
Generalized minimal residual \sep Single-life reinforcement learning \sep Linear algebra \sep Infinite-horizon RL \sep Off-policy learning


\end{keyword}

\end{frontmatter}



\section{Introduction}

Classical iterative methods such as Jacobi, Gauss-Seidel, and Successive Over-Relaxation (SOR) have been used extensively for solving linear systems of equations \cite{Varga1962}. Krylov subspace methods, e.g., Conjugate Gradient (CG) and Bi-Conjugate Gradient (BiCG), work significantly better compared to classical methods due to their faster convergence rates and ability to solve large, sparse linear systems of equations efficiently. Among these methods, Generalized Minimum Residual (GMRES) has been used in commercialized software and research because of its robustness in handling nonsymmetric and large sparse matrices \cite{Saad2003}. GMRES is an iterative Krylov subspace method for solving large and sparse systems of linear equations in the form of  \(Ax = b\)  \cite{SaadSchultz1986,zou2023gmres}. This method is used in a wide range of applications, e.g. fluid mechanics, structural analysis, and finance, to solve sparse and large-scale linear systems derived from the discretization of partial differential equations (PDEs) within complex domains \cite{SaadSchultz1986,ramanathan2003noninvasive}. The GMRES method is often used when direct methods (e.g. Gaussian elimination) are practically challenging to implement or impossible with the available computational power due to the high dimensionality of the matrices governing the physical problem and their \( O(n^3) \) complexity. 

The original GMRES method was developed to minimize the Euclidean norm of the residual vector within the Krylov subspace at each iteration \cite{o1980block,saad1986gmres,golub1977block}. Considering the residual \(r_k = b - Ax_k\) on a Krylov subspace, the cost of iterating on a subspace the same size as matrix \(A\) is considerably high and requires high memory allocation. Therefore, research on how to decrease the number of iterations using preconditioning (e.g., incomplete LU (ILU) and block Jacobi) has been an active area of interest \cite{mittal2003efficient, hegland1992block, anzt2019adaptive}. Restarted GMRES, also known as GMRES($m$), was introduced to limit the dimension of the Krylov subspace to $m$, where $m$ is smaller than the matrix size ($n$) \cite{axelsson1988restarted,joubert1994convergence}. The restart process in GMRES is mainly used to overcome the memory allocation issues associated with full GMRES. In other words, restarts control the storage requirements. In exact arithmetic, GMRES, similar to other Krylov methods, converges within n steps without restarts. However, when n is large, this approach is impractical due to excessive storage and computational costs. The effectiveness of GMRES largely depends on the strategy for selecting when to restart. For a long time, it was assumed that larger restart values are required to mimic the convergence behavior of full GMRES. However, some studies have proposed that smaller restart values are useful to prevent stagnation in convergence \cite{embree2003tortoise}. Mathematical relationships between residual vectors have also been proposed as a method to accelerate convergence in GMRES(m) \cite{baker2009simple,baker2005technique}. In recent years, researchers have integrated machine learning approaches to improve the performance and effectiveness of iterative solvers. RL has shown promise in dynamically adjusting the solver parameters to achieve improved convergence rates \cite{HanLiE2018}. Proximal Policy Optimization (PPO), which is known for its effectiveness and convenient implementation, has been applied effectively to adjust preconditioning in real-time \cite{SchulmanEtAl2017}. Studies carried out on the application of RL to vary the restart parameter in the restarted GMRES method based on the residual norm showed only marginal improvements over existing algorithms. This outcome suggests the need for further research in this area \cite{peairs2011using}.

The objective of this study is to enhance the efficiency of solving linear systems. Unlike physics-informed neural networks (PINNs) and neural operators (NOs), which require partial differential equations (PDEs), this research focuses on developing a faster high-fidelity solver with broader applications, including finance, computer graphics, and image processing \cite{raissi2019physics,kovachki2023neural}. To the best of our knowledge, the impact of the Krylov subspace dimension has not been fully explored in both mathematical and algorithmic contexts. This study investigates the use of RL to guide the choice of Krylov subspace dimension. Using the residual vectors obtained from each iteration as the state representation of the environment, a single-life RL off-policy agent is trained to achieve faster convergence rates and solve larger matrices with the available computational resources. This agent is integrated with the solver in an online setting and operates without requiring pre-training using an SLRL approach \cite{chen2022you}.

\section {Background}
GMRES is an iterative method for solving large systems of linear equations of the form:
\begin{equation}
    A\mathbf{x} = \mathbf{b} 
\end{equation}

where \( A \in \mathbb{R}^{n \times n} \) is the coefficient matrix, \( \mathbf{x} \in \mathbb{R}^n \) is the unknown solution vector, and \( \mathbf{b} \in \mathbb{R}^n \) is the right-hand side vector. These systems appear in most scientific and engineering applications from the discretization of PDEs, machine learning, and optimization. At each iteration, GMRES finds the vector \( \mathbf{x}_m \) within the Krylov subspace $K_m(A, \mathbf{r}_0)$ that minimizes the Euclidean norm of the residual \( \| \mathbf{b} - A\mathbf{x}_m \|_2 \). The Krylov subspace of dimension $m$ is defined as:

\begin{equation} \label{kryloveq}
     K_m(A, \mathbf{r}_0) = \text{span} \{ \mathbf{r}_0, A\mathbf{r}_0, A^2\mathbf{r}_0, \ldots, A^{m-1}\mathbf{r}_0 \} 
\end{equation}

where \( \mathbf{r}_0 = \mathbf{b} - A\mathbf{x}_0 \) is the initial residual value corresponding to an initial guess \( \mathbf{x}_0 \). In order to prevent numerical instability in iterative process and efficiency in the storage and computation, an orthonormal basis of the Krylov subspace is constructed using the Arnoldi iteration. Orthonormalization starts by normalizing the vector \(\mathbf{r}_0\), to obtain the first basis vector \( \mathbf{v}_1 = \mathbf{r}_0 / \| \mathbf{r}_0 \| \), in a sequence to generate vectors \( \mathbf{v}_2, \mathbf{v}_3, \ldots, \mathbf{v}_m \) to construct the columns of an orthonormal matrix \( V_m = [\mathbf{v}_1, \mathbf{v}_2, \ldots, \mathbf{v}_m] \). The process continues to generate an upper Hessenberg matrix \( H_m \in \mathbb{R}^{(m+1) \times m} \) with the following structure:

\[
H_m = \begin{pmatrix}
h_{11} & h_{12} & h_{13} & \cdots & h_{1m} \\
h_{21} & h_{22} & h_{23} & \cdots & h_{2m} \\
0 & h_{32} & h_{33} & \cdots & h_{3m} \\
\vdots & \vdots & \vdots & \ddots & \vdots \\
0 & 0 & 0 & \cdots & h_{m+1,m}
\end{pmatrix}.
\]

In the GMRES algorithm, the approximate solution \( \mathbf{x}_m \) is expressed as:

\[
\mathbf{x}_m = \mathbf{x}_0 + V_m \mathbf{y},
\]

in which \( \mathbf{y} \in \mathbb{R}^m \) is determined by solving the least-squares problem:

\[
\min_{\mathbf{y} \in \mathbb{R}^m} \left\| \beta \mathbf{e}_1 - H_m \mathbf{y} \right\|,
\]

with \( \beta = \| \mathbf{r}_0 \| \) and \( \mathbf{e}_1 \in \mathbb{R}^{m+1} \) being the first canonical basis vector.

In GMRES(m), after \(m\) iterations, GMRES restarts with the new residual \(r_m\) and constructs a new Krylov subspace \(K_m(A, r_m)\). This process repeats until the residual norm is reduced below a specified tolerance. Bellow are the key matrices involved in GMRES(m) that can convert the search space from  \(n\) dimension to \(m\) where $m \leq n$.

\begin{enumerate}
    \item \textbf{Matrix \(A\)}:
    \[
    A = \begin{pmatrix}
    a_{11} & a_{12} & \cdots & a_{1n} \\
    a_{21} & a_{22} & \cdots & a_{2n} \\
    \vdots & \vdots & \ddots & \vdots \\
    a_{n1} & a_{n2} & \cdots & a_{nn}
    \end{pmatrix}
    \]
    \item \textbf{Orthonormal Basis \(V_m\)}:
    \[
    V_m = \begin{pmatrix}
    v_{11} & v_{12} & \cdots & v_{1m} \\
    v_{21} & v_{22} & \cdots & v_{2m} \\
    \vdots & \vdots & \ddots & \vdots \\
    v_{n1} & v_{n2} & \cdots & v_{nm}
    \end{pmatrix}
    \]
    \item \textbf{Upper Hessenberg Matrix \( H_m \)}:
    \[
    H_m = \begin{pmatrix}
    h_{11} & h_{12} & \cdots & h_{1m} \\
    h_{21} & h_{22} & \cdots & h_{2m} \\
    0 & h_{32} & \cdots & h_{3m} \\
    \vdots & \vdots & \ddots & \vdots \\
    0 & 0 & \cdots & h_{m+1,m}
    \end{pmatrix}.
    \]
\end{enumerate}

\section{Methodology}

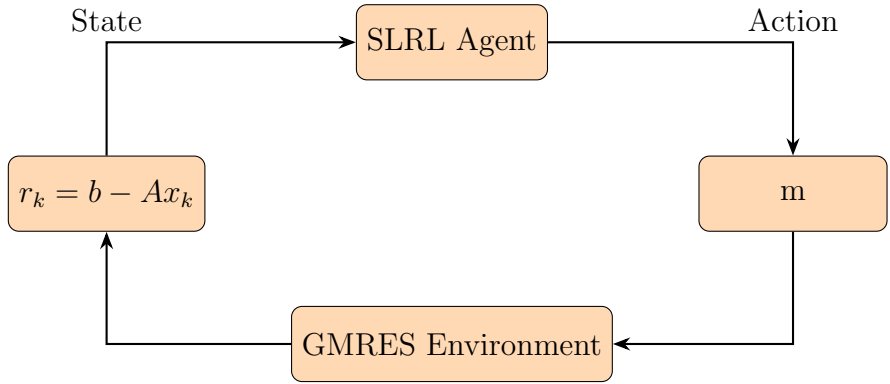
\begin{figure}[h]
    \centering
    \begin{tikzpicture}[
        box/.style={rectangle, draw, fill=orange!30, minimum width=2.5cm, minimum height=1cm},
        rounded/.style={rectangle, rounded corners, draw, fill=orange!30, minimum width=2.5cm, minimum height=1cm},
        arrow/.style={-Stealth, thick}
        ]

        \node[rounded] (agent) {SLRL Agent};
 
        \node[rounded, below right=1 cm and 2cm of agent] (below_right_box) {m};
        \node[rounded, below left =1 cm and 2cm of agent] (below_left_box) {\(r_k = b - Ax_k\) };
        \node[rounded, below= 3 cm of agent] (environment) { GMRES Environment};


        \draw[arrow] (agent.east) -| node[above] {Action}(below_right_box.north);
        \draw[arrow] (below_right_box.south) |- (environment.east);
        \draw[arrow] (environment.west) -| (below_left_box.south);
        \draw[arrow] (below_left_box.north) |- node[above] {State}(agent.west);

    \end{tikzpicture}

    \caption{SLRL architecture for Adaptive changes in Krylov subspace dimension}
    \label{fig:rl_diagram}
\end{figure}

An SLRL agent is trained to learn and adjust the restart parameter (\( m \)) based on the current residual vector and the change in residuals. In episodic reinforcement learning, it is assumed that resets occur at regular intervals, often every few hundred or thousand timesteps. Episodic RL is not practical for iterative solvers, since the primary goal of solvers is to minimize the residual as efficiently as possible to achieve convergence. In contrast to episodic RL, which requires training the agent, SLRL operates without requiring pre-training and resetting. In SLRL the agent lives as long as the task is available. In our case, the agent will live as long as the Euclidean norm of the residual is higher than the specified tolerance. The schematics of the proposed framework is shown in Figure \ref{fig:rl_diagram}. The state \( s_t \) at time \( t \) is represented by the residual vector \( r_k \). The action \( a_t \) is the restart parameter \( m \) for the next iteration. The reward \( R_t \) is defined to be inversely correlated with the residual norm added by the improvement in the residual norm for each iteration, \( R_t =   ( cte / \|r_k\| ) + (\|r_{k-1}\| - \|r_k\|)\) to encourage actions that lead to faster reductions in residuals. It is important to note that this residual is different from the relative residual reported by SciPy. The state is calculated as \(r_k = b - Ax_k\), but the residual plotted in this study is the relative residual that is consistent with the callback from GMRES in SciPy. A soft actor-critical agent (SAC) is trained to interact with a GMRES(m) solver environment to learn the optimal policy $\pi^*$ that maps states to actions. The pseudocode for the AK-SLRL algorithm is shown in Algorithm \ref{Algo}. Since the reward function is strictly increasing with each interaction in the environment, a time penalty is considered in the reward function. This approach is used so that the agent is not satisfied with the increase in rewards and takes actions that further improve the rewards. By introducing a time penalty to the agent, the agent is encouraged to optimize its actions to compensate for the losses due to the time limits.

SAC is an off-policy learning algorithm that provides sample efficiency, which is critical to reducing the number of restarts. We used a short replay buffer and a discount factor ($\gamma$ of 0.97) for efficient learning of the residual history. Observation and action values are mapped and normalized to the range [0,1]. Although actions are integers between 1 and $m$, they are mapped to the range [0,1]. A single episode approach is used for the learning algorithm. The number of timesteps per episode is large enough to ensure convergence within each episode. The high rate of exploration is used at the beginning of the episode. This high exploration rate prevents the agent from becoming trapped in suboptimal policies that could result from strictly increasing reward. As discussed above, residuals generally decrease with each environment interaction, which results in higher rewards. Therefore, a high level of initial exploration is necessary for the agent to continue searching for better actions that enhance the reward and prevent premature convergence to suboptimal solutions. A short replay buffer ( half the size of the matrix with a maximum size of 20,000 ) is used. This can save the memory and accelerate the solver. This short replay buffer is used based on studies that suggested the relation between residual vector in the last three vectors can improve the convergence of GMRES(m) solver \cite{ayachour2003fast}.

\begin{algorithm}[H]
\caption{Adaptive Krylov Subspace Exploration with SLRL}\label{Algo}
\begin{algorithmic}[1]
    \State \textbf{Initialize:}
    \State \quad Set initial guess $x_0$.
    \State \quad Compute initial residual $r_0 = b - A x_0$.
    
    \State \textbf{Set Initial State:} $s_0 = r_0$.
    
    \While{not converged \textbf{or} $k < \texttt{max\_iter}$}
        \State \textbf{Action:}
        \State \quad $a_t \in [0, 1] \rightarrow m \in \mathbb{N}$ \quad \text{(maximum GMRES restart)}
        \State \quad 
        \[
            m = \left\lfloor a_t \times (m_{\text{max}} - 1) \right\rfloor + 1
        \]
        
        \State \textbf{Arnoldi Iteration:} Perform Arnoldi iteration for $m$ steps to construct $V_m$ and $H_m$.
        \For{$j = 1$ \textbf{to} $m$}
            \State Compute $w = A v_j$.
            \State Orthogonalize $w$ against $v_1, v_2, \ldots, v_j$:
            \State \quad $h_{ij} = v_i^\top w$ \quad \text{for} $i = 1, 2, \ldots, j$
            \State \quad $w = w - \sum_{i=1}^{j} h_{ij} v_i$
            \State Normalize $w$ to obtain $v_{j+1}$:
            \State \quad $h_{j+1,j} = \| w \|$
            \State \quad $v_{j+1} = \frac{w}{h_{j+1,j}}$
        \EndFor
        
        \State \textbf{Solve Least-Squares Problem:}
        \State \quad Solve $\min_{y \in \mathbb{R}^m} \| \beta e_1 - H_m y \|$ to obtain $y_m$.
        
        \State \textbf{Update Solution:}
        \State \quad $x_{t+1} = x_t + V_m y_m$
        
        \State \textbf{Compute New Residual:}
        \State \quad $r_{k+1} = b - A x_{t+1}$
        
        \State \textbf{Update State:}
        \State \quad $s_{t+1} = r_{k+1}$
        
        \State \textbf{Compute Reward:}
        \[
            R_t = \left( \frac{\text{cte}}{\| r_t \|} \right) + \left( \| r_{t-1} \| - \| r_t \| \right)
        \]
        
        \State \textbf{Training:}
        \State \quad Update the SAC agent according to the reward $R_t$.
    \EndWhile
    
    \If{$\| r_{k+1} \|$ is below a specified tolerance}
        \State \textbf{Terminate.}
    \Else
        \State \textbf{Repeat the iteration.}
    \EndIf
\end{algorithmic}
\end{algorithm}

\section{Results and Discussions}

A selection of matrices from the SuiteSparse Matrix Collection is used to assess the performance of the AK-SLRL algorithm \cite{davis2011university}. To ensure solver convergence with restarted GMRES, the search is restricted to positive definite matrices with more than 90\% symmetry. A total of 110 matrices with dimensions greater than 1,000, are evaluated. The dimension of the Krylov subspace is constrained to 20 in all experiments. 

The results show that when the solver requires more than 1,000 iterations, AK-SLRL achieves convergence 5 to 30 times faster than constant restart. The convergence rate depends on the size of the matrix, condition number, and symmetry. To demonstrate the convergence behavior of the AK-SLRL agent, a diverse set of matrices is selected from various application domains, as presented in Table \ref{tab:selected_matrices_convergence} \cite{davis2011university}. Although we used 110 matrices to test the robustness of the proposed algorithm, a section of them is shown in Table \ref{tab:selected_matrices_convergence} to visualize the residuals and spot the differences. This table lists the application area, sizes, and number of nonzero elements for the matrices.
\begin{table}[ht]
    \centering
    \caption{Selected Matrices for Visualizing Convergence in Various Application Areas}
    \vspace{0.3cm} 
    \footnotesize
    \begin{tabular}{lllrr}
        \toprule
        Problem Name & Application Area & Size ($n$) & Nonzeros (nnz) \\
        \midrule
        1138\_bus           & Power Network                 & 1,138    & 4,054      \\
        Finance256          & Finance portfolio optimization           & 37,376   & 298,496   \\
        Ct20stif        & Structural finite elements  & 52,329  & 2,600,295  \\
        Olesnik0            & Mining system modeling                & 88,263   & 744,216    \\
        ex19                & Computational Fluid Dynamics  & 12,005   & 259,577    \\
        Crankseg\_1   & Structural finite elements  & 52,804   & 10,614,210    \\
        \bottomrule
    \end{tabular}
    \label{tab:selected_matrices_convergence}
\end{table}

\begin{figure}[H]
    \centering
    \subfigure[$1138\_bus$]{
        \includegraphics[width=0.46\textwidth]{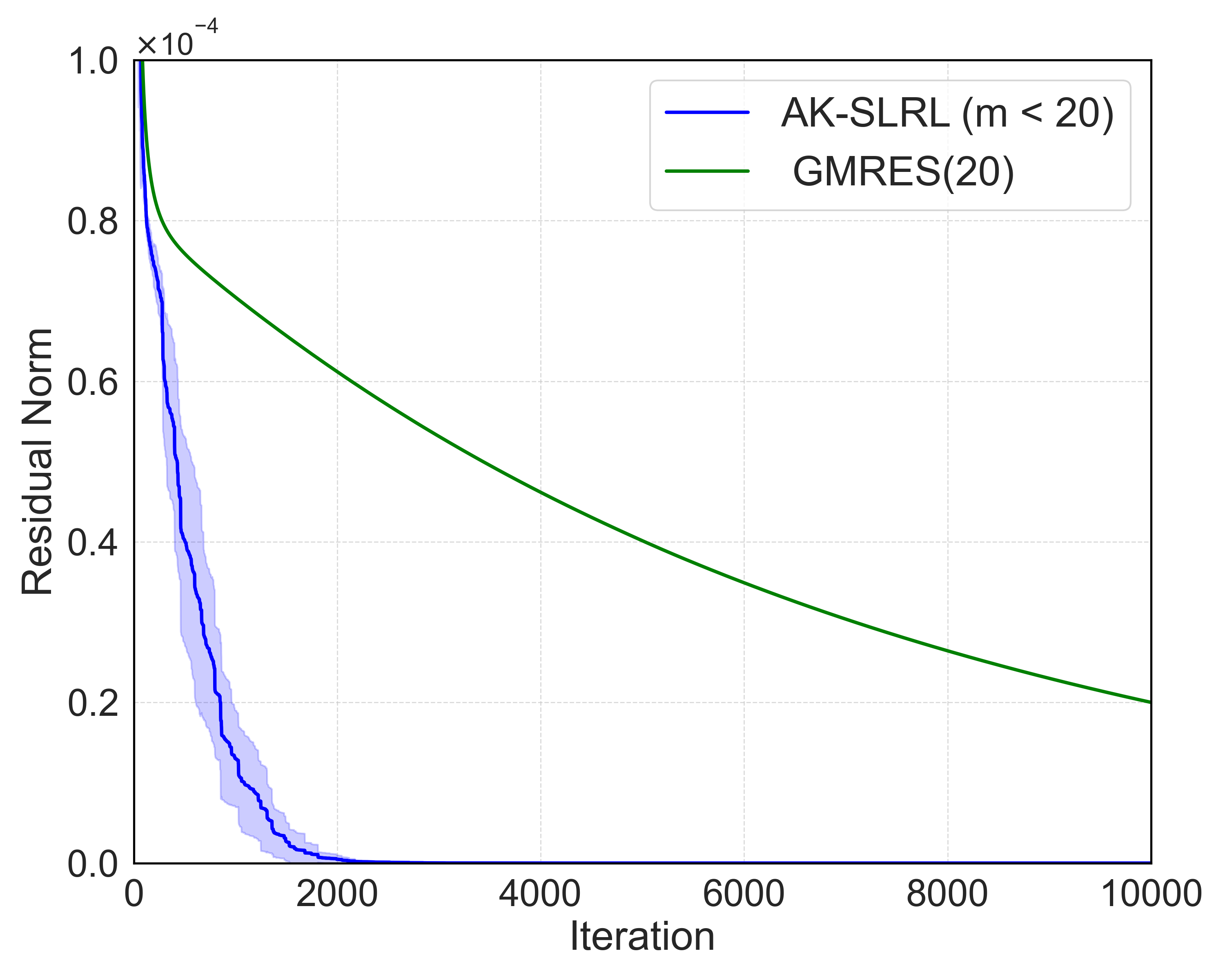}
    }
    \subfigure[Finance256]{
        \includegraphics[width=0.46\textwidth]{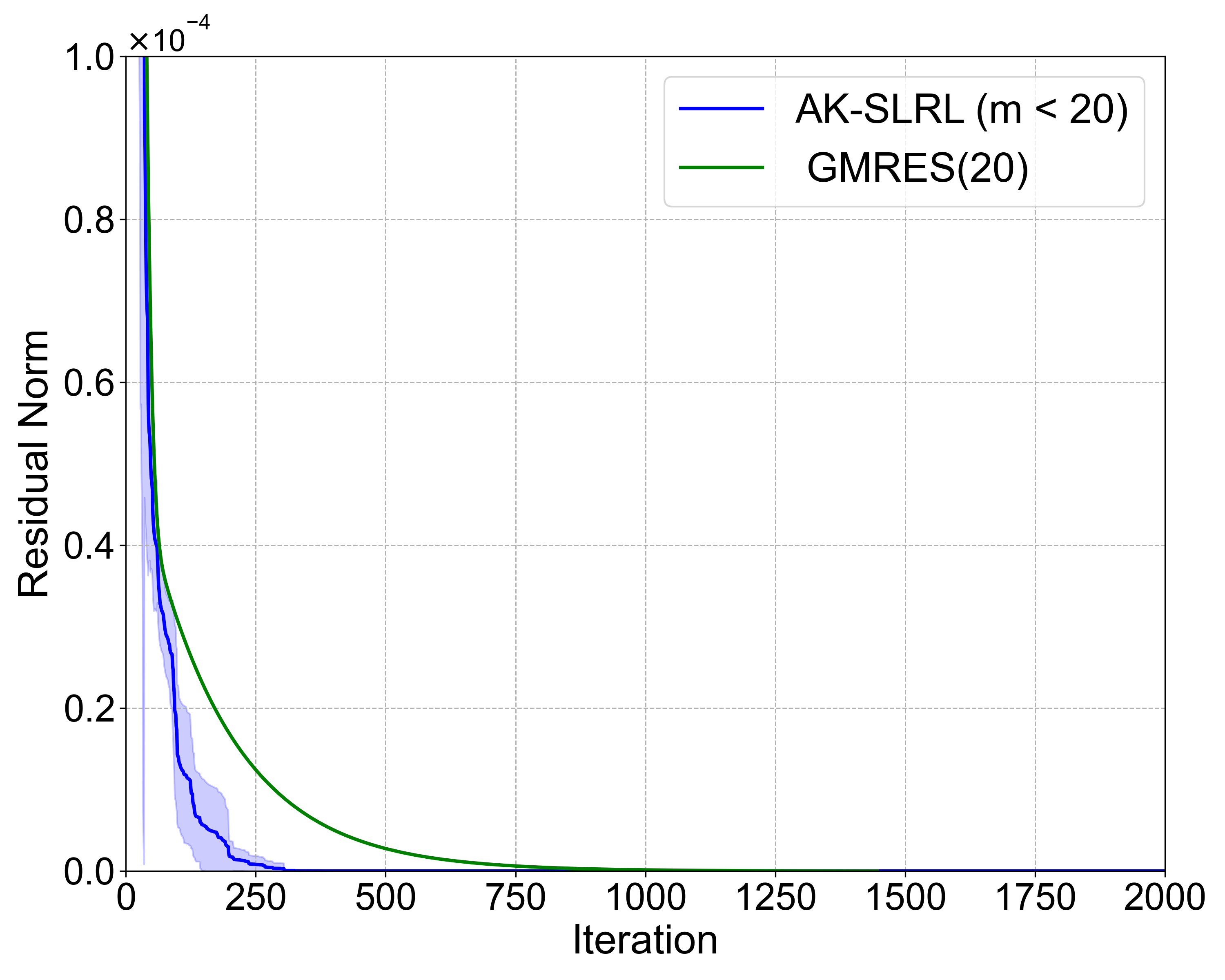}
    }
    \subfigure[Ct20stif]{
        \includegraphics[width=0.46\textwidth]{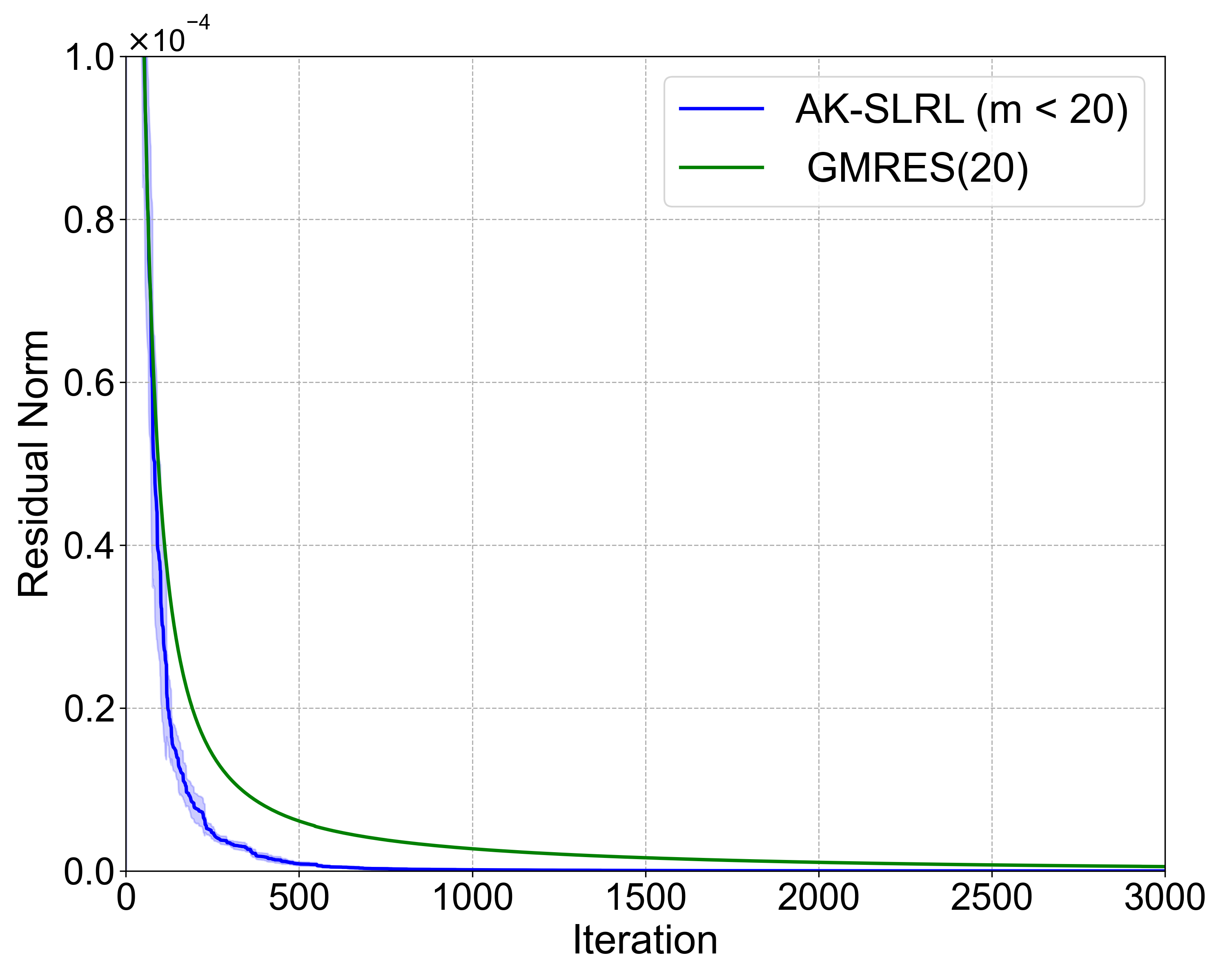}
    }
    \subfigure[Olesnik0]{
        \includegraphics[width=0.46\textwidth]{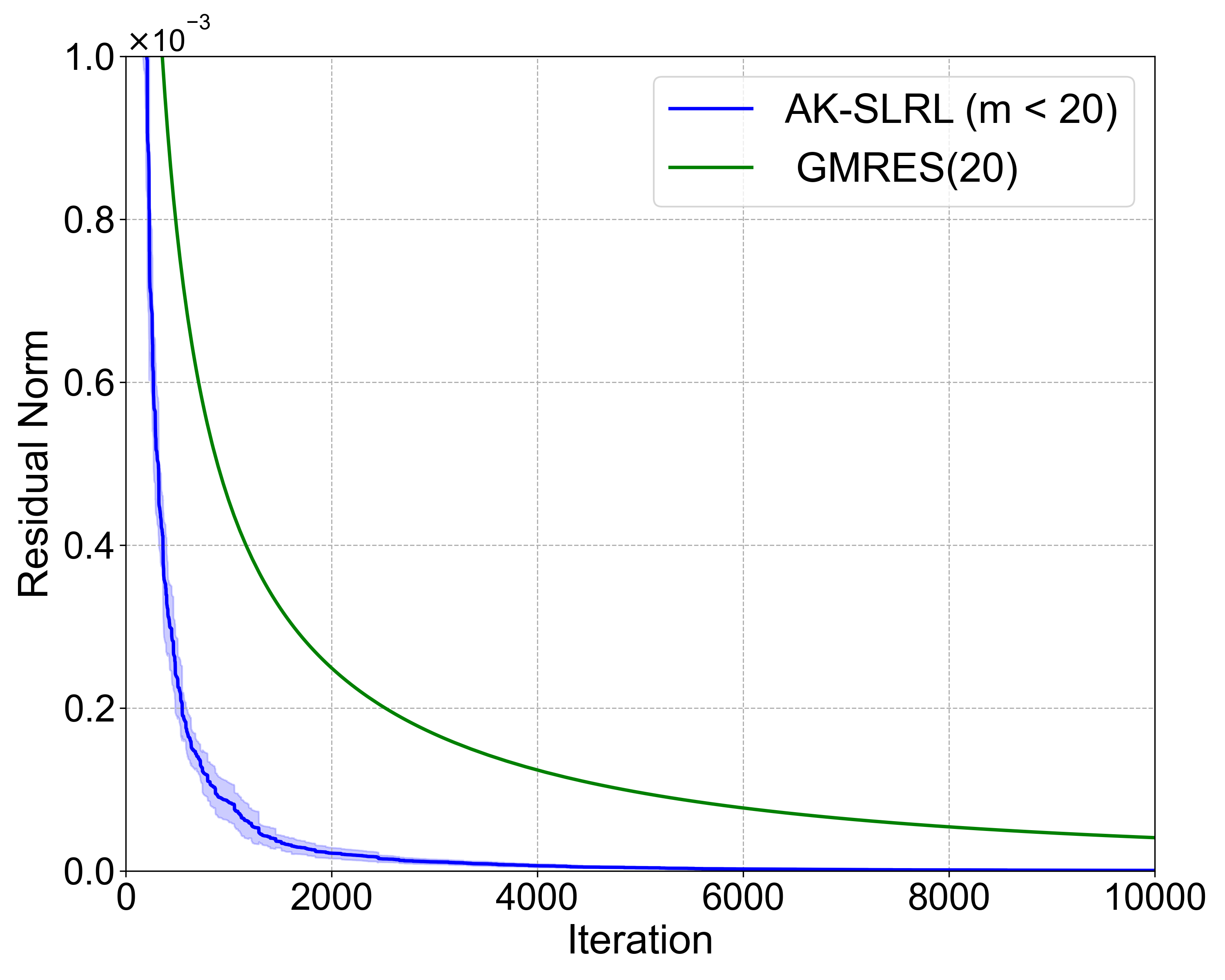}
    }
    \subfigure[Ex19]{
        \includegraphics[width=0.46\textwidth]{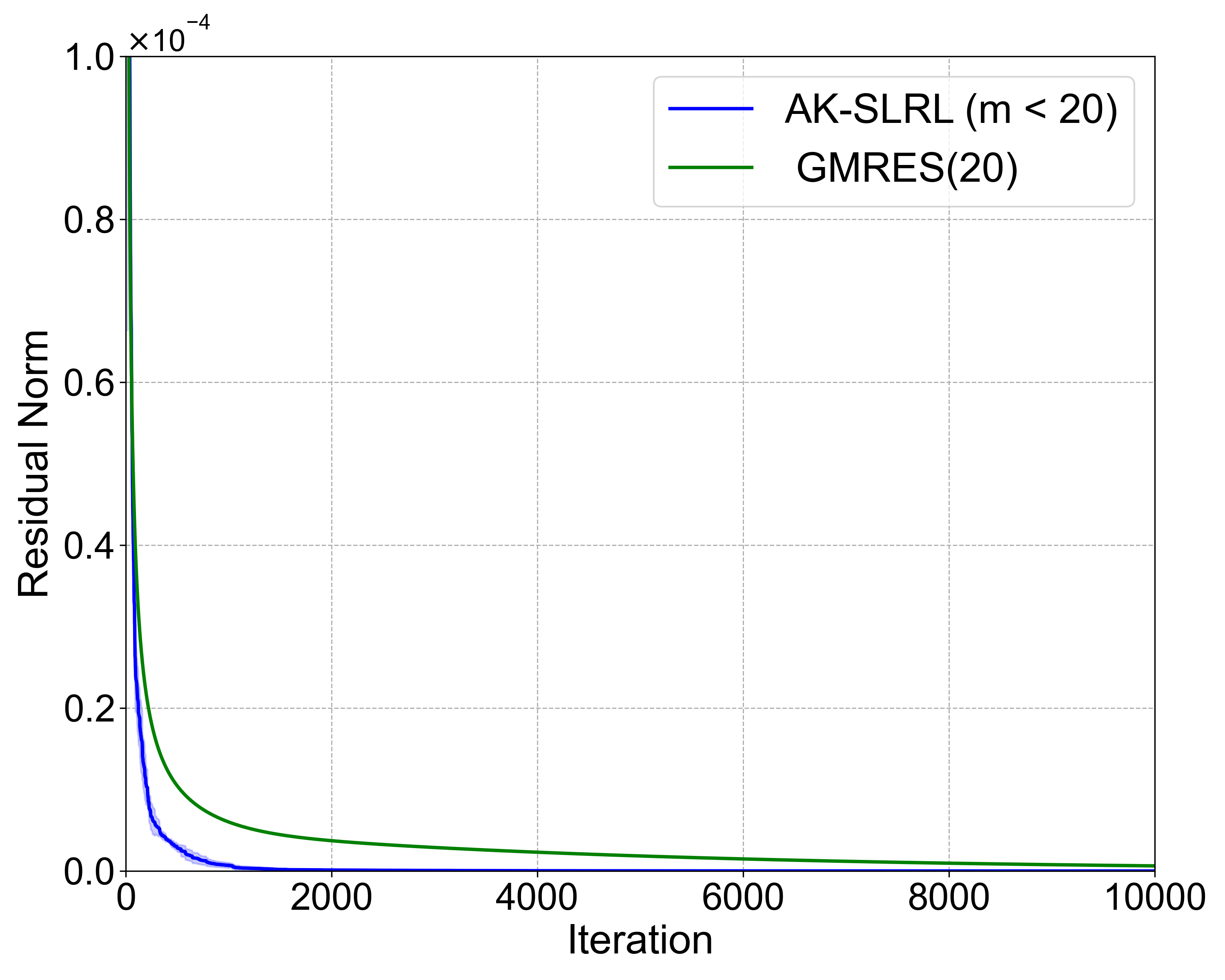}
    }
    \subfigure[$Crankseg\_1$]{
        \includegraphics[width=0.46\textwidth]{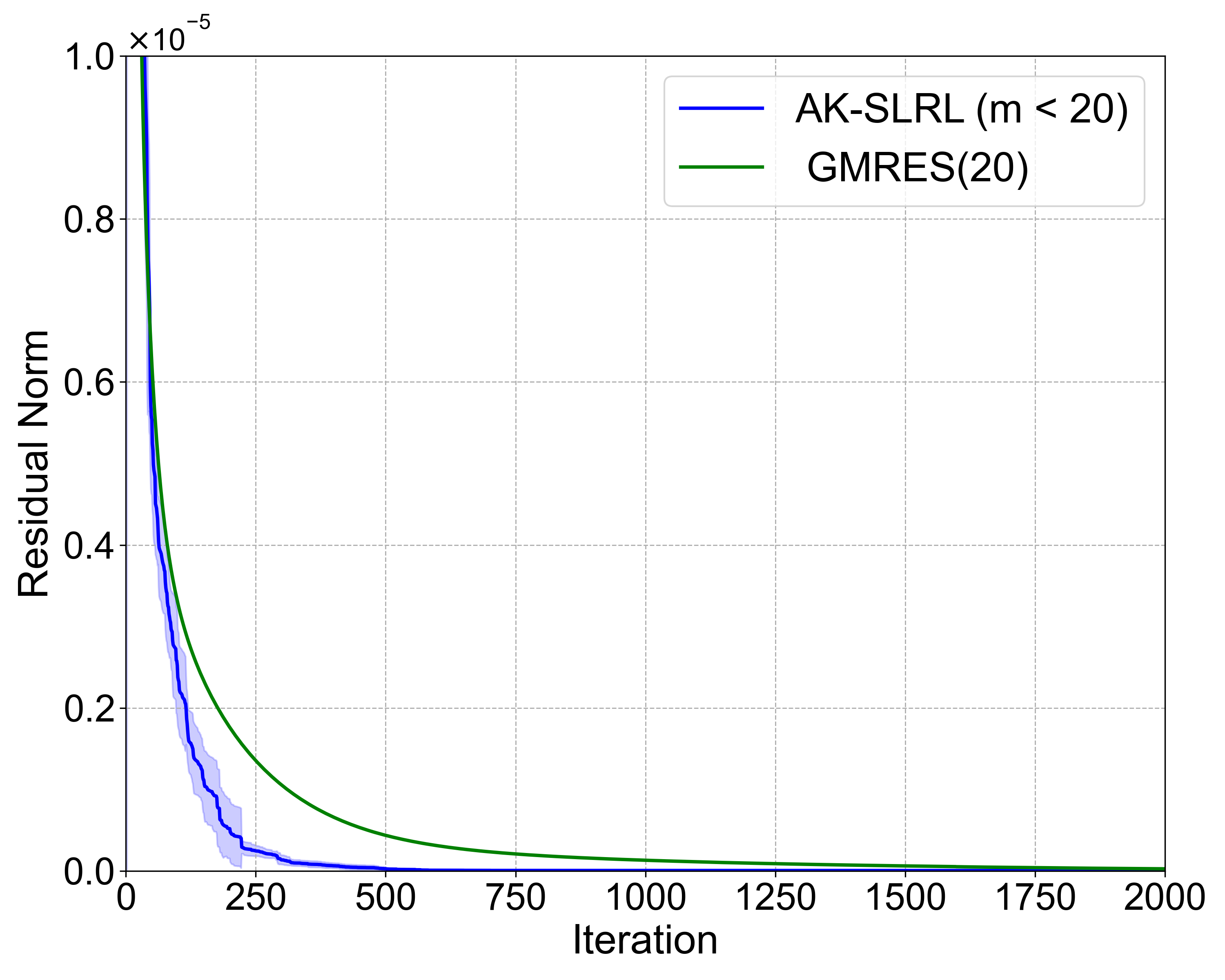}
    }
    \caption{Comparison of the Euclidean norm of the residuals for the constant restart GMRES and AK-SLRL algorithms}
    \label{residuals}
\end{figure}

The Euclidean norm of the residuals for the constant restart GMRES and AK-SLRL algorithms for the matrices of Table \ref{tab:selected_matrices_convergence} are shown in Figure \ref{residuals}. The residuals reported here are the mean of 10 experimental runs along with the standard deviation represented by the shaded areas. It can be seen that the AK-SLRL algorithm converges significantly faster in all cases. A notable observation is the substantial reduction in residuals occurring after a few hundred or one thousand iterations. In cases where GMRES with constant restart struggle with low progression in residual reduction or local stagnation, AK-SLRL learns how to skip the stagnation and converge faster. This is particularly helpful in extremely large matrices where increasing the dimension of the Krylov subspace is not feasible. In this context, an iteration refers to one cycle, which contains \( m \)  Arnoldi steps.

\begin{figure}[H]
    \centering
    \subfigure[$1138\_bus$]{
        \includegraphics[width=0.46\textwidth]{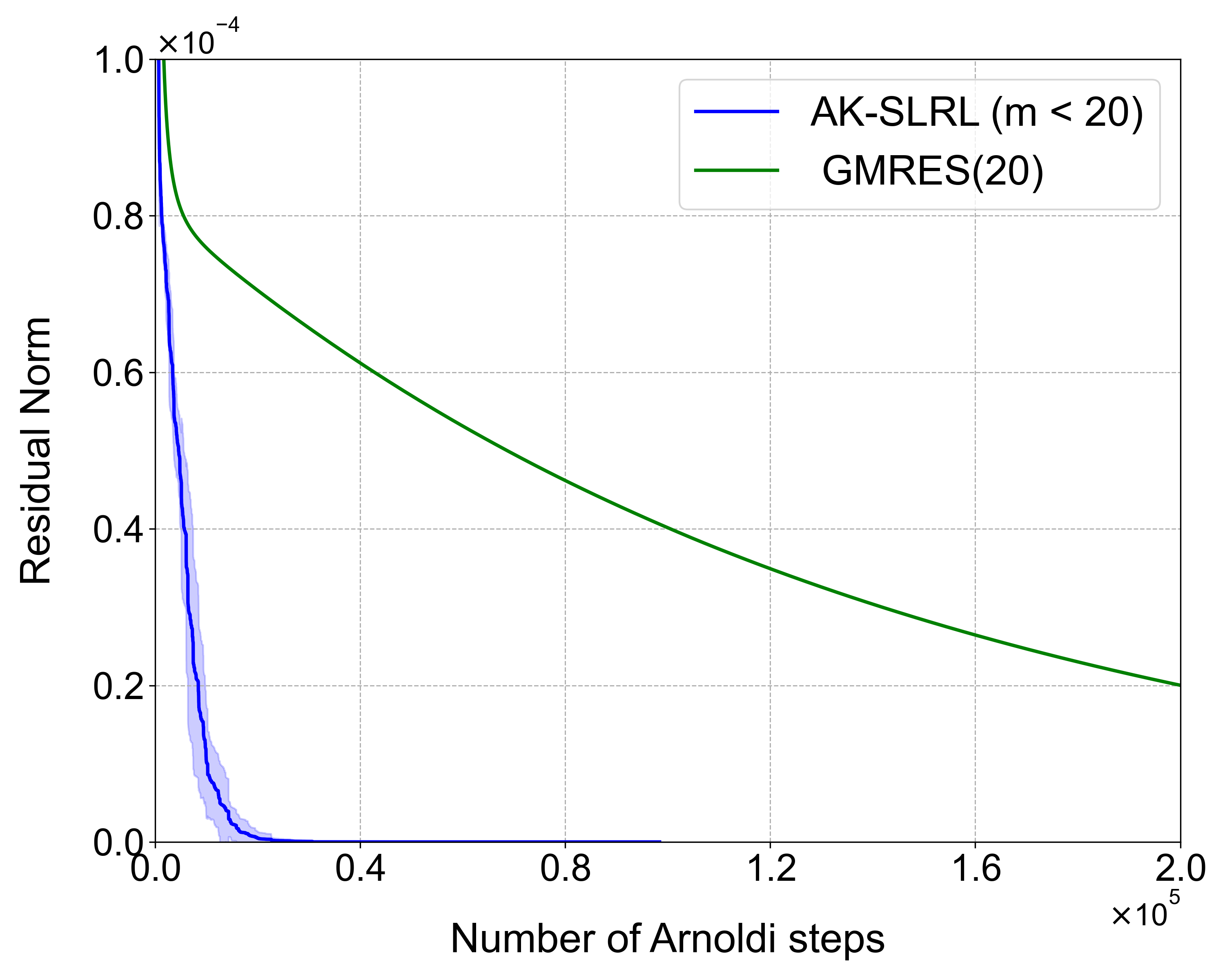}
    }
    \subfigure[Olesnik0]{
        \includegraphics[width=0.46\textwidth]{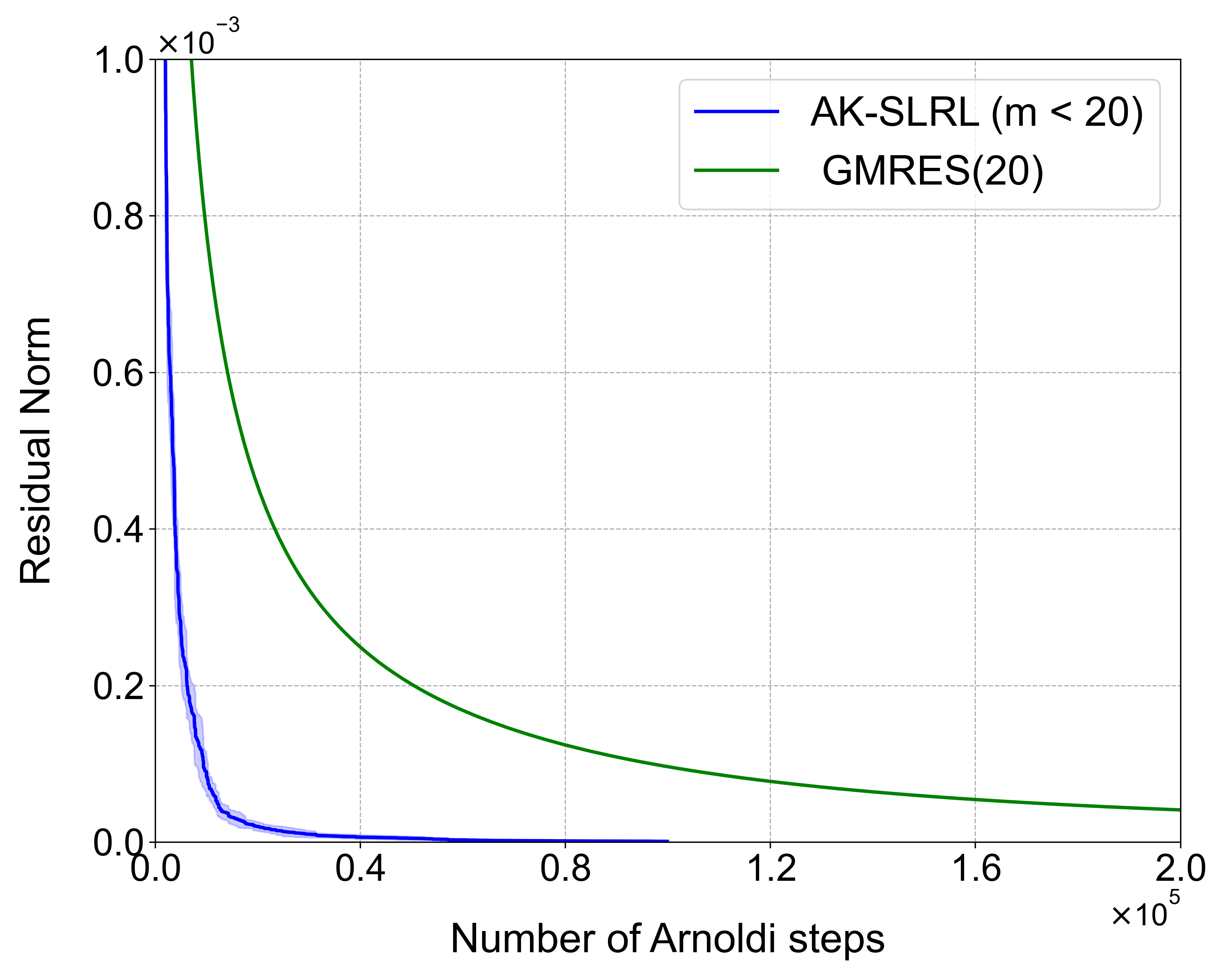}
    }
    \caption{Comparison of residual curves versus the number of Arnoldi steps taken for each method}
    \label{fig:ArnoldiStep}
\end{figure}

For residual plots, it is more appropriate to use the number of Arnoldi steps as the horizontal axis rather than the number of cycles. This is because each Arnoldi step involves a single application of the matrix-vector product, and the least-squares computation is proportional to the number of Arnoldi steps, which corresponds to the size of the upper Hessenberg matrix. The number of Arnoldi steps taken in the AK-SLRL algorithm is significantly smaller than in GMRES(20) because the value of \( m \) in AK-SLRL is smaller than 20 in each iteration. In this study, we did not use any preconditioner to facilitate a direct comparison between the methods. However, it is worth noting that when using a preconditioner, one additional application of the preconditioner is required per Arnoldi step. This further increases the computational cost of the GMRES method for each Arnoldi step, regardless of whether it includes restarting or not. To facilitate demonstration and further show the advantage of AK-SLRL, we used the number of Arnoldi iterations on the horizontal axis in Figure \ref{fig:ArnoldiStep}. This figure shows the number of Arnoldi steps performed for convergence when using the AK-SLRL algorithm compared to GMRES (20) for two different matrix from Table \ref{tab:selected_matrices_convergence}. The difference in convergence becomes clearer when comparing the number of Arnoldi iterations. For example, when solving the power network problem represented by matrix $1138\_bus$, the AK-SLRL method converges with 20 times fewer Arnoldi iterations than GMRES(20). 

In general, GMRES(m) is preferred over full GMRES for large problems due to its significantly lower memory requirements while maintaining reasonable computational efficiency. In cases where sufficient memory is available and rapid convergence is critical, full GMRES might offer faster performance. However, for most commercial computers used in scientific computing, memory constraints are typically more limiting than processor speed. Therefore, AK-SLRL is crucial for  efficiency in solving large-scale problems.

\begin{figure}[H]
    \centering
    \subfigure[First 50 iterations]{
        \includegraphics[width=0.46\textwidth]{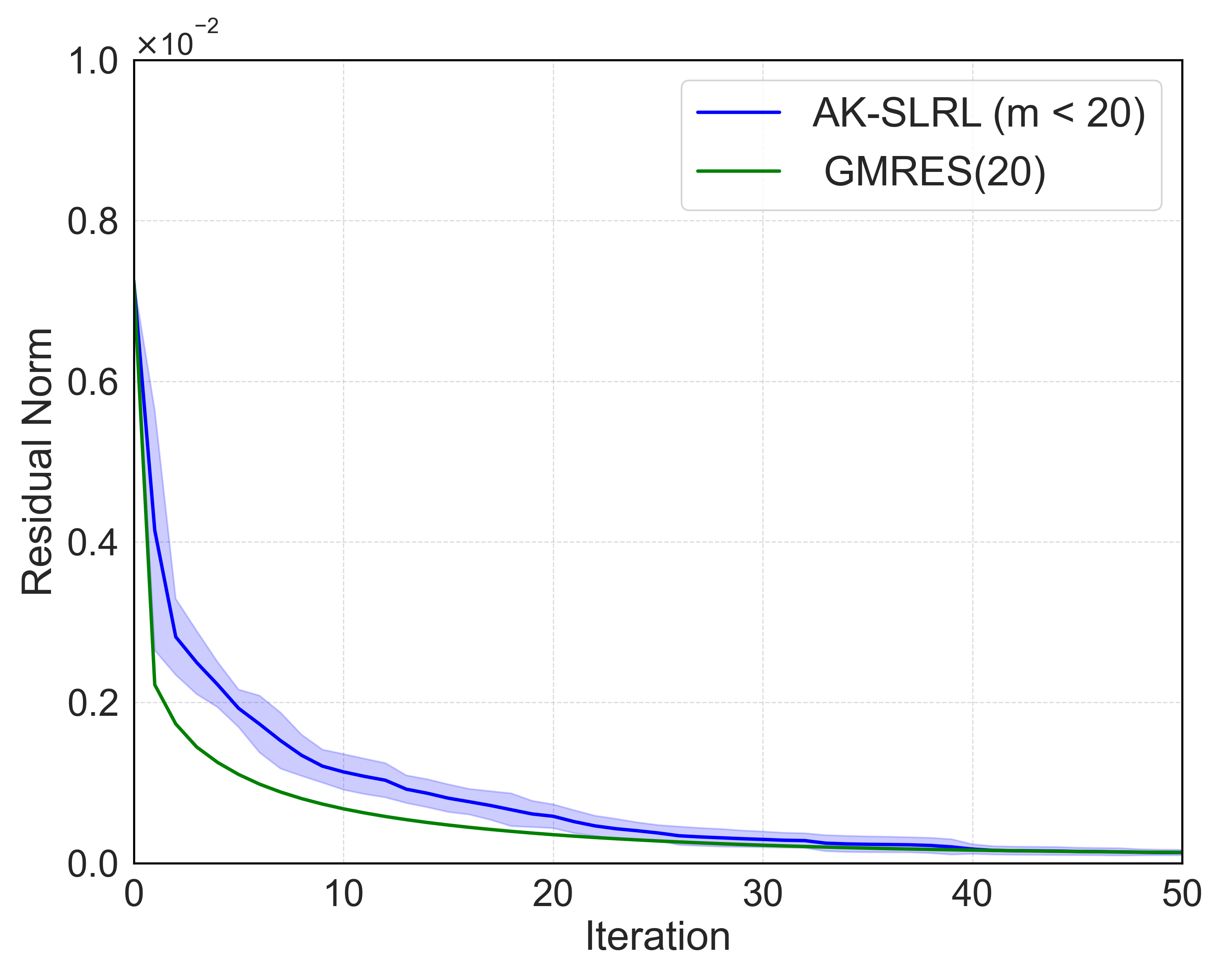}
    }
    \subfigure[First 200 iterations]{
        \includegraphics[width=0.46\textwidth]{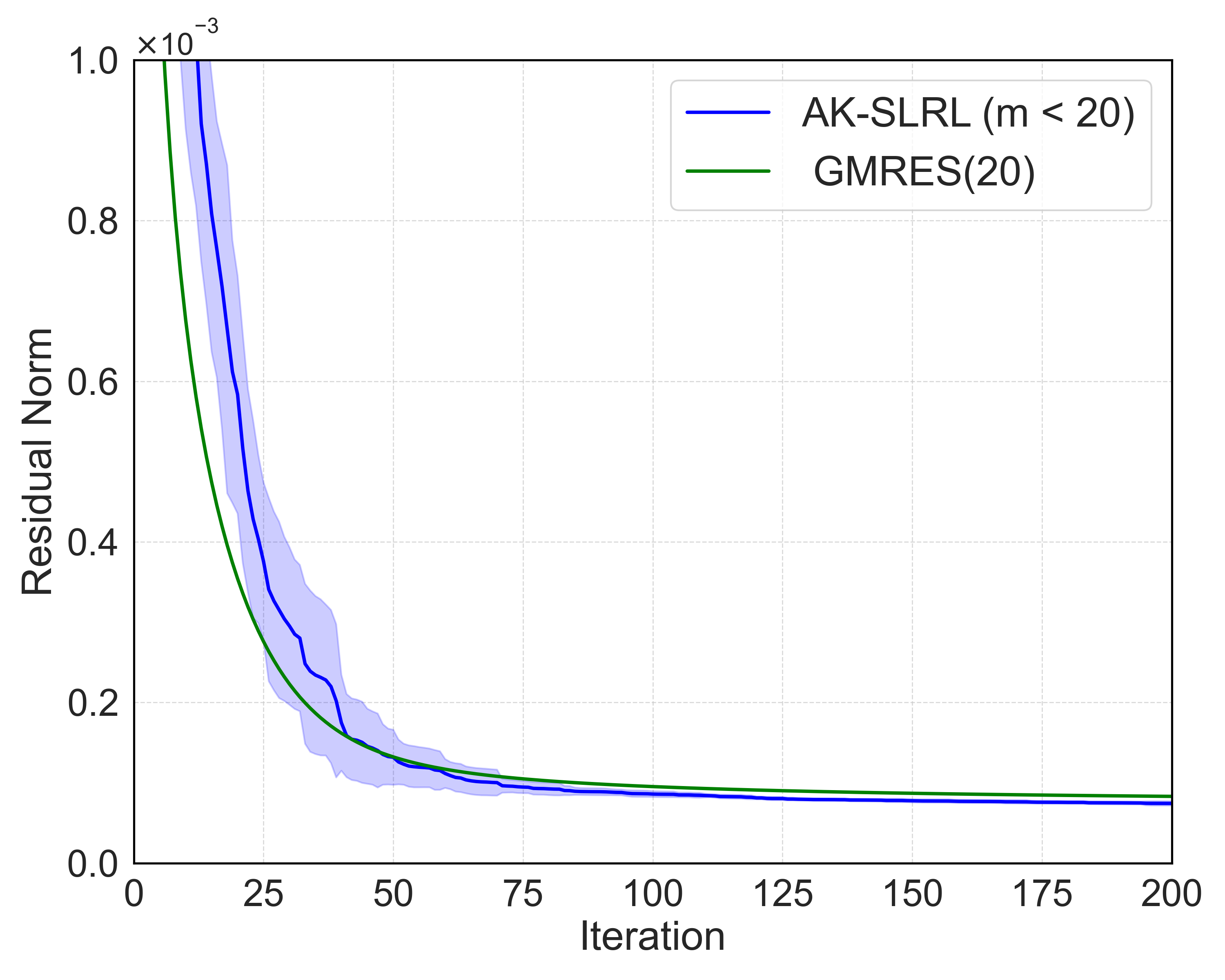}
    }
    \caption{Residual curves for the initial iterations of the matrix for power network problem ($1138\_bus$)}
    \label{fig:Initialsteps}
\end{figure}

Figure \ref{fig:Initialsteps} shows the initial residual history for the power network matrix ($1138\_bus$). As can be seen in the graphs, the constant Krylov dimension has less residual at the beginning of the life of the AK-SLRL (Figure \ref{fig:Initialsteps}(a)). However, after few hundreds iterations, the agent learns to increase the reward by reducing the residuals (Figure \ref{fig:Initialsteps}(b)) and reduce the time required for reaching the solution (Figure \ref{residuals}(a)).

For all symmetric positive definite (SPD) matrices, the RL-based solver presents a significant speed-up. For matrices with a larger number of nonzero elements, the GMRES(m) method experiences a slowdown in convergence. The RL-based solver, however, reduces the residual norm at each iteration and makes progress towards convergence with much lower residual values.

Considering that a short-memory experience replay buffer is used, the RL-based solver can have a positive impact on solving PDEs where memory is limited due to the large and sparse matrix size. The reduction in computation time is significant due to the \(\mathcal{O}(m^2)\) relationship between the dimension of the Krylov subspace and the computation cost. Reducing the dimension of the Krylov subspace not only requires less memory but also reduces the cost of the orthogonalization process and the cost of solving the least-squares problem. In theory, if \(m\) is reduced by half in GMRES(m), the computational cost per outer iteration decreases by 75\%, assuming a similar effectiveness in reducing the residual. As shown in the plots, SLRL can significantly enhance the effectiveness of solving linear systems of equations, resulting in more than a 20× faster convergence for most cases, depending on the matrix size and condition.

\section{Conclusion}

In this study, an adaptive approach for dimensioning the Krylov subspace using SLRL is introduced. The adaptive change in the dimension of the Krylov subspace using SLRL improves the convergence of the GMRES iterative solver by up to 30×. Single-life RL significantly accelerates both the convergence and computation time of the GMRES solver by reducing the size of the Krylov subspace in each iteration while improving the residuals toward faster convergence.

Learning from residual vectors and a short replay buffer, the SLRL agent takes actions that increase the speed of the linear solver by 5 to 30 times. This approach represents a considerable step towards achieving a fast linear solver using an off-policy agent that lives only once. The results can have a significant impact on a wide range of applications that require solving linear systems of equations, including physics simulations, finance, and computer graphics. With a specified memory allocation, it is possible to solve larger matrices more quickly without pre-training using AK-SLRL. This opens the door to faster computation and the ability to solve more complex problems.

In future research, the authors plan to study preconditioning techniques that take advantage of information derived from Arnoldi iterations. The Arnoldi process provides valuable information on the transition from the dominant eigenspace to smaller eigenspaces. This information can be captured using graph-structured data to construct the initial guess and preconditioner for the GMRES algorithms.

 \bibliographystyle{elsarticle-num} 
 \bibliography{elsarticle-template-num}




\end{document}